\documentstyle[15lomcon,cite,epsfig]{article}

\begin{document}

\title{Searches for Supersymmetry with the CMS Experiment}

\author{ Altan Cakir \footnote{e-mail: cakir@cern.ch},
on behalf of the CMS Collaboration}

\address{Deutsches Elektronen-Synchrotron, 22607, Hamburg, Germany}


\maketitle\abstracts{After a very successful startup of the LHC in 2010, the CMS experiment has already accumulated significantly more data in 2011. After the successful re-discovery of the Standard Model, the search for signs of new physics has already reached, and in most cases enlarged, the limits from previous experiments. In this conference report I review the recent discovery reach of SUSY searches that will be performed with the 2011 data.}

\section{Introduction}
\label{sec:intro}

High energy and particle physics enter a new era with the start-up of the Large Hadron Collider (LHC) at CERN, a proton-proton accelerator located at the Swiss-French border with a circumference of 27 km. The LHC provides a new energy regime to explore the origin of the electroweak symmetry breaking, search for and study Supersymmetry (SUSY) or other Beyond the Standard Model (BSM) physics scenarios. 

Physics beyond the SM is expected to consist of new heavy particles, otherwise these particles would have been discovered already at previous accelerators. These heavy particles decay to lighter particles, which will have higher transverse momentum ($P_{T}$) with respect to the beam axis than decay products of light particles. In addition, SUSY is expected to produce missing transverse energy ($E^{Miss}_{T}$) due to the escaping the lightest supersymmetric particle (LSP), neutralinos\footnote[2]{Assuming R-parity, R=$ (-1)^{3(B-L)+2S}$ where B and L are baryon- and lepton numbers, respectively and S is the spin, is conserved. All SM particles have even R-parity, while SUSY particles are R-odd.},   candidate for dark matter\cite{martin}. At $\sqrt{s}$ = $7$ TeV center-of-mass energy, the production cross-sections may be sufficient such that a data sample of modest integrated luminosity, $1.1 fb^{-1}$ could contain a large number of new particles. 

The Compact Muon Solenoid (CMS) is an experiment \cite{cms} designed to find evidence for new physics beyond SM using a signature of high-energy objects in the final state. The signatures expected for new physics have been taken into consideration extensively in the design of the experiment. In this report various SUSY analyses based on signatures involving jets, leptons, and missing transverse momentum are discussed. These analyses are designed to be simple and generic, focusing on basic topological and kinematic properties that typically characterize SUSY signatures at the LHC.

This conference report is organized as follows: the search for a missing energy signature in di-jet and multi-jet events using the $\alpha_{T}$ variable is discussed for Hadronic SUSY searches. In the following sub-section two analyses with leptonic signature, one requiring two same-charge leptons and the other requiring opposite-charge leptons are presented, respectively. To interpret the results, a practical model of SUSY breaking, the constrained minimal SUSY model (cMSSM \footnote[3]{The cMSSM is described by five parameters: the universal scalar and gaugino mass parameters ($m_0$ and $m_{1/2}$, respectively), the universal trilinear soft SUSY breaking parameter $A_0$, and two low-energy parameters, the ratio of the two vacuum expectation values of the two Higgs doublets, tan$\beta$, and the sign of the Higgs mixing parameter, sign($\mu$). The parameter values defining LM$4$ are $210, 285, 0, 10, +$ and LM$6$ are $85, 400, 0, 10, +$, respectively.}) is discussed.

\section{SUSY Searches}
\label{sec:SUSY}
\subsection{Results for Hadronic SUSY Searches with $\alpha_{T}$ parameter}
\label{sec:HSUSY}
The kinematic variable $\alpha_{T}$ was first defined for di-jet \cite{alphaTh} and subsequently extended to n-jet events  \cite{alphaT}. In hadronic SUSY searches it is based on the assumption that colored particle (squark-squark, gluino-gluino, or gluino-squark) are pair-produced and subsequently decay directly to a quark and neutralino. The extension to n-jets allows sensitivity to beyond the direct decay. This approach is the most promising for points in a SUSY parameter space where squarks have large branching ratios to decay directly to the LSP.

\begin{figure}[hbtp]
\centering
\includegraphics[height=4cm]{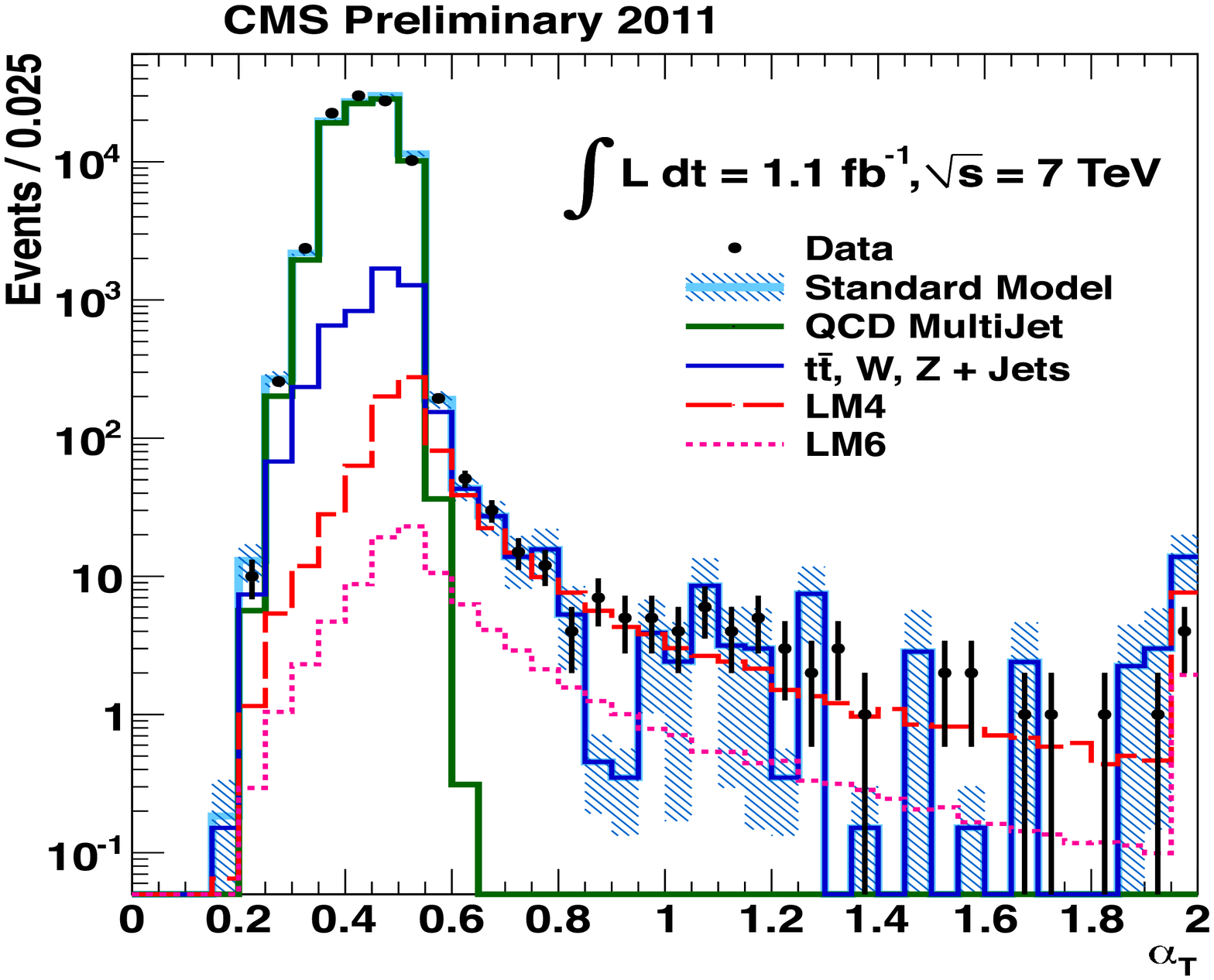}
 \includegraphics[height=4cm]{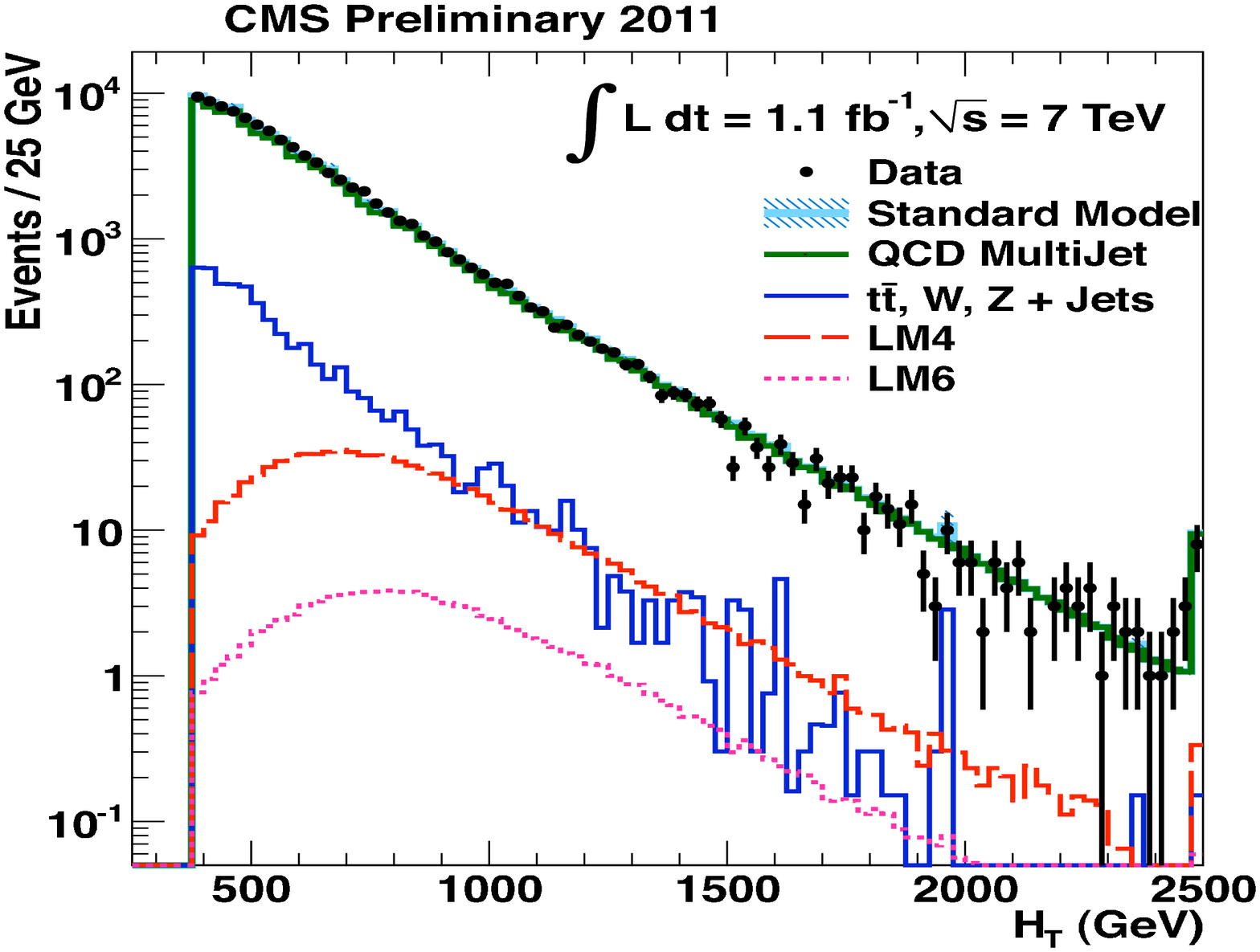}
\caption{Comparison of the $\alpha_{T}$ (left) and $H_T$ (right) distribution between data and MC.}
\end{figure}

The data sample used in this analysis is recorded with a trigger based on the scalar sum of the transverse energy $E_T$ of jets, defined in general as $H_T$ = $\Sigma^{Njet}_{i=1} E^{j_{i}}_{T}$  , where Njet is the number of jets. Events are selected if they satisfy $H^{Trigger}_{T} > 375$ GeV. The expected event topology consists of two or more high $p_T$ jets and two invisible neutralinos leading to a missing transverse energy signature. The main background processes for this topology are QCD di-jet events and Z+jets where Z decays into two invisible neutrinos. Figure $1$ demonstrates that  the $\alpha_{T}$  variable is  an excellent discriminator between QCD background and signal. Though the Monte-Carlo (MC) simulation describes the data reasonably well, the background yields are obtained from data control samples. The ratio $R_{\alpha_{T}}$ = $N_{\alpha_{T}>\theta}/N_{\alpha_{T}<\theta}$ shows no dependence on $H_T$ if $\theta>0.55$ is chosen such that the numerator of the ratio in all $H_T$ bins is dominated by $t \bar{t}$, W +jets and Z$\rightarrow \nu \bar{\nu}$ + jets events. Figure $2$ shows that data are consistent with SM by predictions. Data-driven techniques are also used to estimate the $t \bar{t}$ by W$\rightarrow$l$\nu$, and the Z background from $\gamma$+jets events. 

\begin{figure}[hbtp]
\centering
\includegraphics[height=4cm]{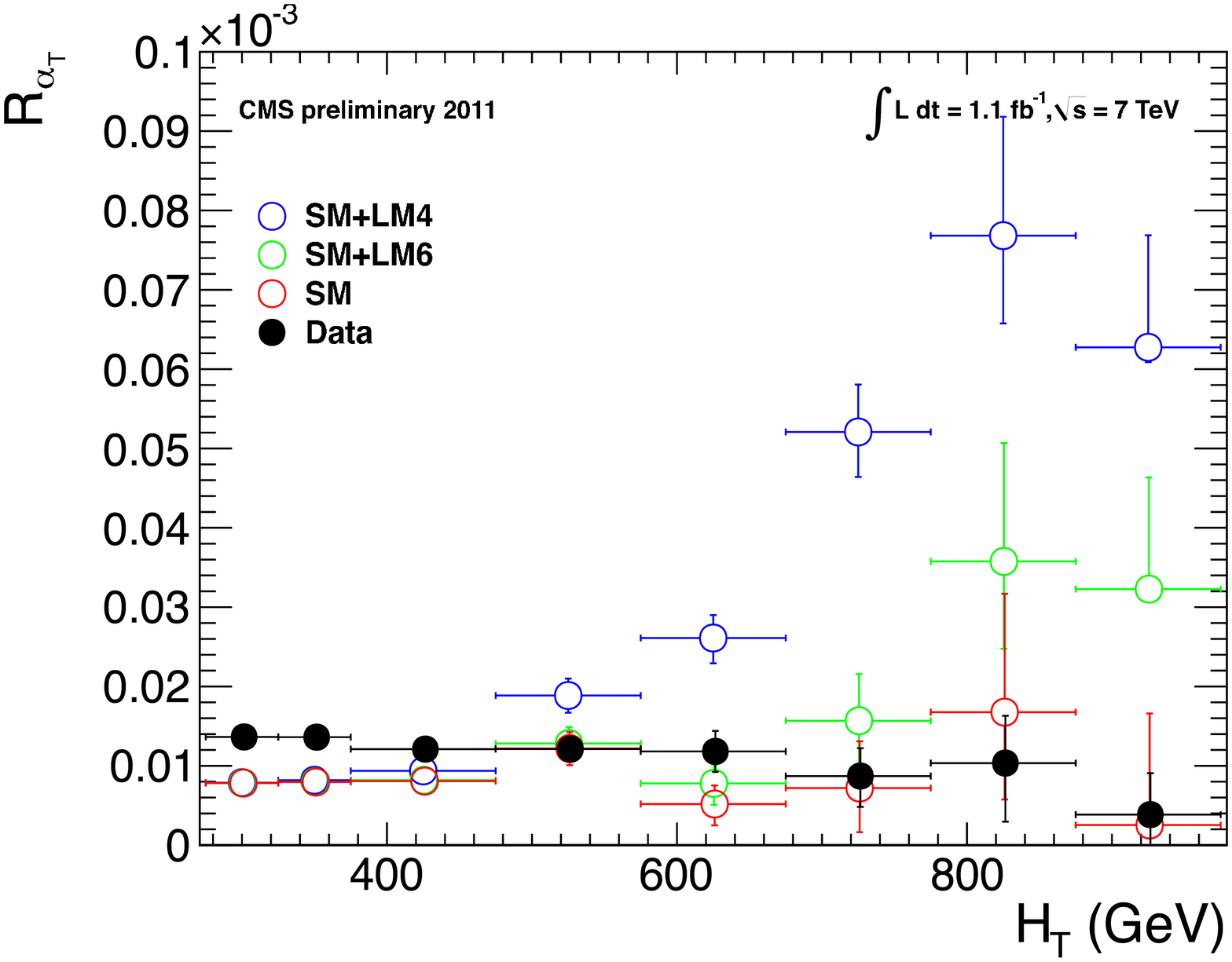}
 \includegraphics[height=4cm]{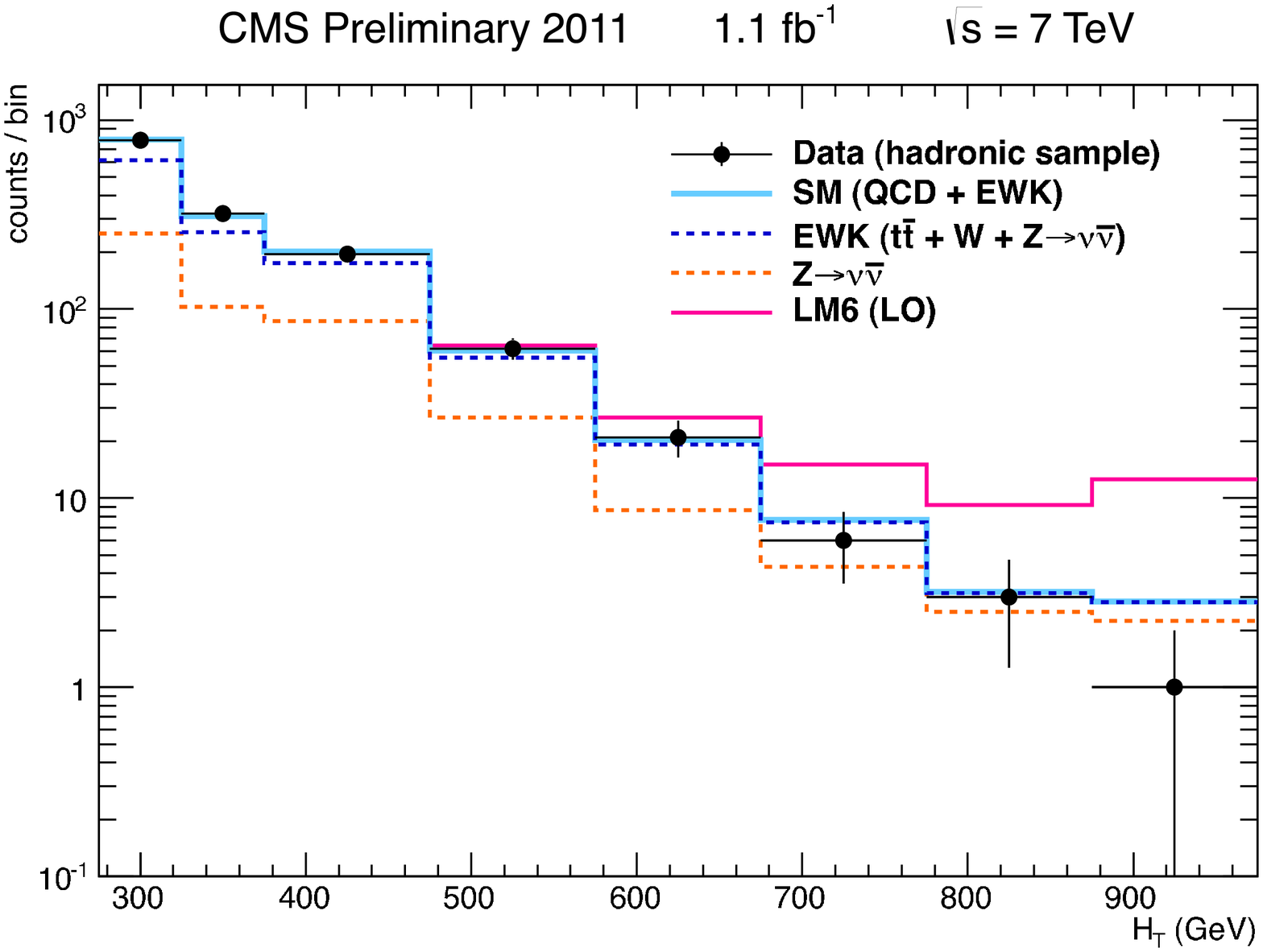}
\caption{The dependence of $R_{\alpha_{T}}$ (left) on $H_T$ for events with $N_{jets} \ge 2$. $H_T$ (right) distribution for the events observed in data, the outcome of the fit (light blue line) and a breakdown the individual background contributions as predicted by the control samples.}
\end{figure}

The results are currently based on an integrated luminosity of $1.1 fb^{-1}$.  The observed yield is consistent with the predictions from MC and from the data-driven background estimates, so that no evidence for SUSY signal is observed.

\subsection{Results for Leptonic SUSY Searches}
\label{sec:LSUSY}
\subsubsection{Dileptonic events with Same Charge }

The requirement of same-sign, isolated, dileptons, as the Standard Model backgrounds are naturally expected to be small, makes this a very appealing channel to study  \cite{SS}. Muon, electron, and tau candidates with $p_T$ as low as $5, 10$, and $15$ GeV respectively, and with
$\mid$$\eta$$\mid$ $< 2.4$, are used to define the dilepton final states. All events considered for search regions are required to have two leptons with the same charge, at least two jets, and $E^{Miss}_{T}$ above $30$ GeV. The requirement of at least two jets provides a universal requirement of $H_T > 80$ GeV. The following selection cuts for four different search regions are defined:
\begin{enumerate}
\item  The requirement of $H_T > 400$ GeV and $E^{Miss}_{T} > 120$ GeV, provides a high
sensitivity to the low values of $m_0$, as in LM6.
\item The requirement of  $H_T > 200$ GeV and $E^{Miss}_{T} > 50$ GeV, provides a high sensitivity to mass-splittings between gluinos/squarks and charginos/neutralinos.
\item The requirement of  $H_T > 400$ GeV and $E^{Miss}_{T} > 50$ GeV, provides a high sensitivity the high values of $m_0$.
\item The requirement of  $H_T > 80$ GeV and $E^{Miss}_{T} > 100$ GeV, provides a high sensitivity to models predicting low hadronic activity with a high $E^{Miss}_{T}$.
\end{enumerate}

\begin{figure}[t]
\centering
\includegraphics[height=4cm]{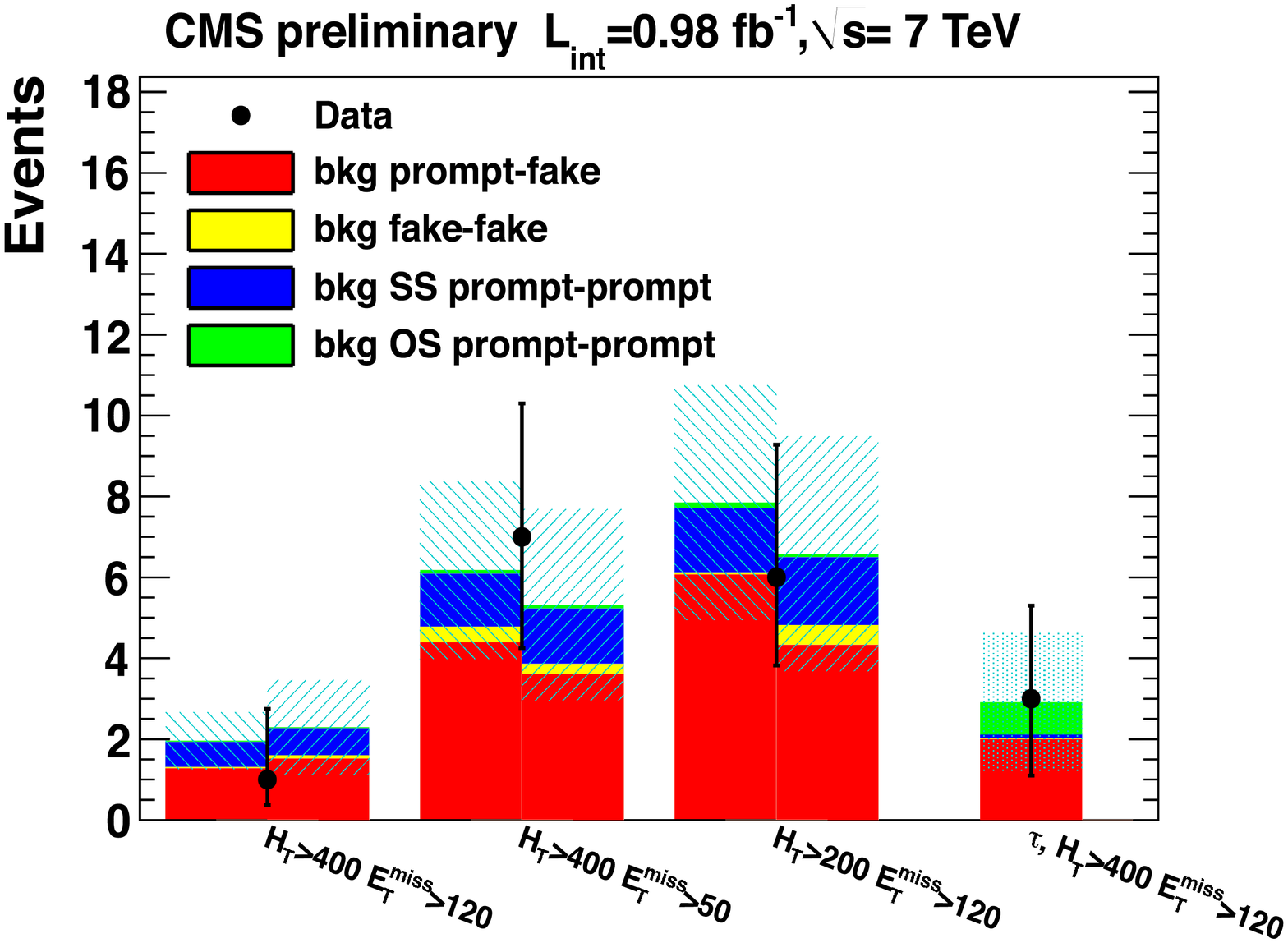}
\includegraphics[height=4cm]{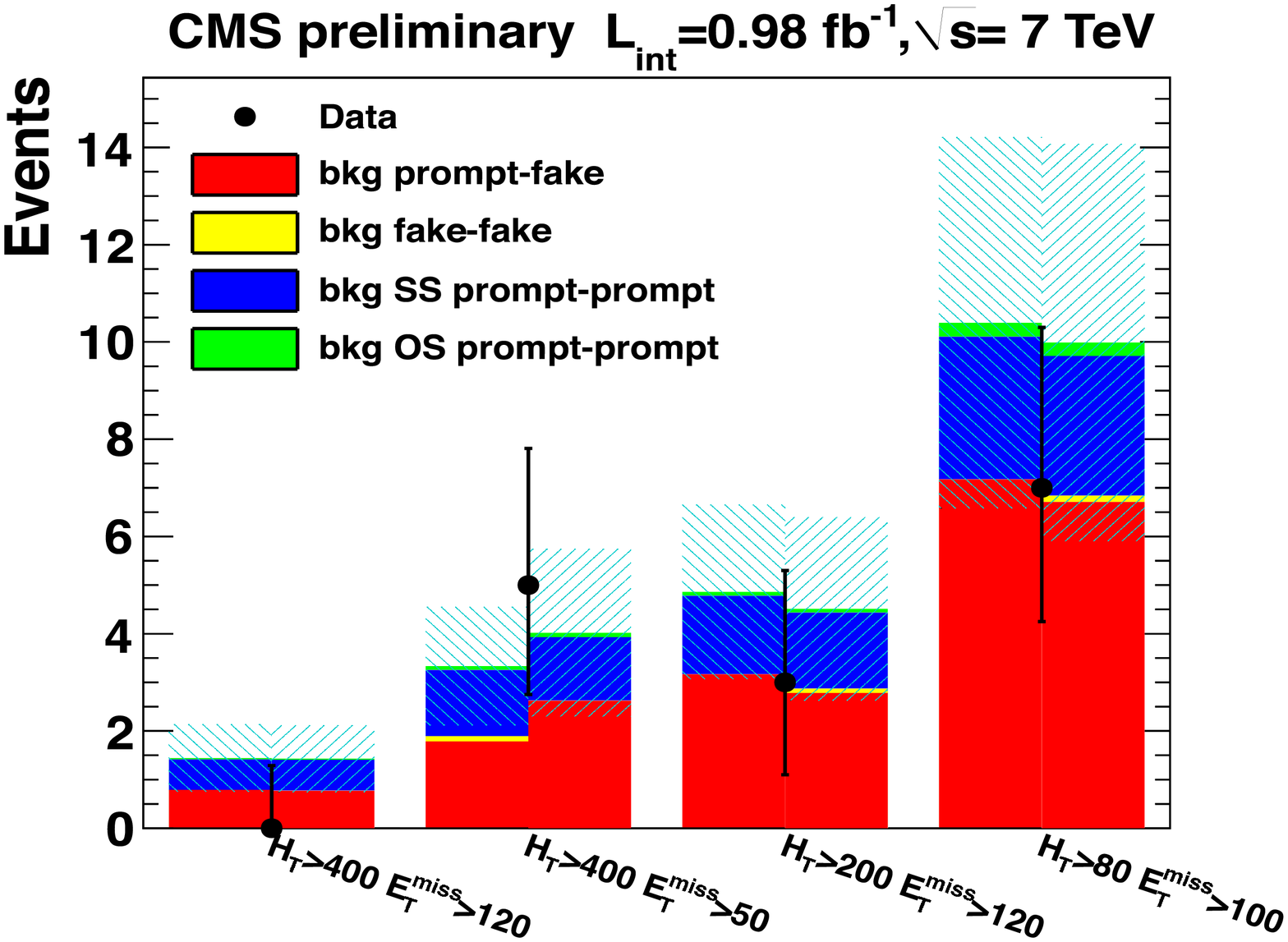}
\caption{Summary of background predictions and observed yields in the search regions for the inclusive (left) selections, dilepton candidates with $H_T>200$ GeV,  and and high-$p_T$ dilepton selections (right), dilepton candidates with both leptons having $p_T>10$ GeV, at least one lepton having $p_T>20$ GeV and no $H_T$ requirement beyond $H_T > 80$ GeV.}
\end{figure}

Figure $3$ summaries the result of searches for new physics with same-sign dilepton events in the ee, $\mu \mu$, e$\mu$, e$\tau$, $\mu \tau$, and $\tau \tau$ final states. No evidence for an excess over the background prediction has been seen at L=$0.98 fb^{-1}$. 

\subsubsection{Dileptonic Events with Opposite Charge}

This analysis focuses on events with opposite charge leptons pairs  ($e^{+} e^{-}$, $e^{+}  \mu^{-}$, $e^{-}  \mu^{+}$, $\mu^{+}  \mu^{-}$) jets and $E^{Miss}_{T}$ in the final state \cite{OS}. Due to the decay $\chi^{0}_{2} \rightarrow \tilde{\ell} \tilde{\ell}$$\rightarrow \chi^{0}_{1} \ell^{\pm} \ell^{\pm}$ in the cMSSM, a characteristic kinematic edge is expected in the invariant dilepton mass distributions. In the following two signal regions are defined by motivating requirements of large $E^{Miss}_{T}$ and $H_T$:

\begin{figure}[t]
\centering
\includegraphics[height=4cm]{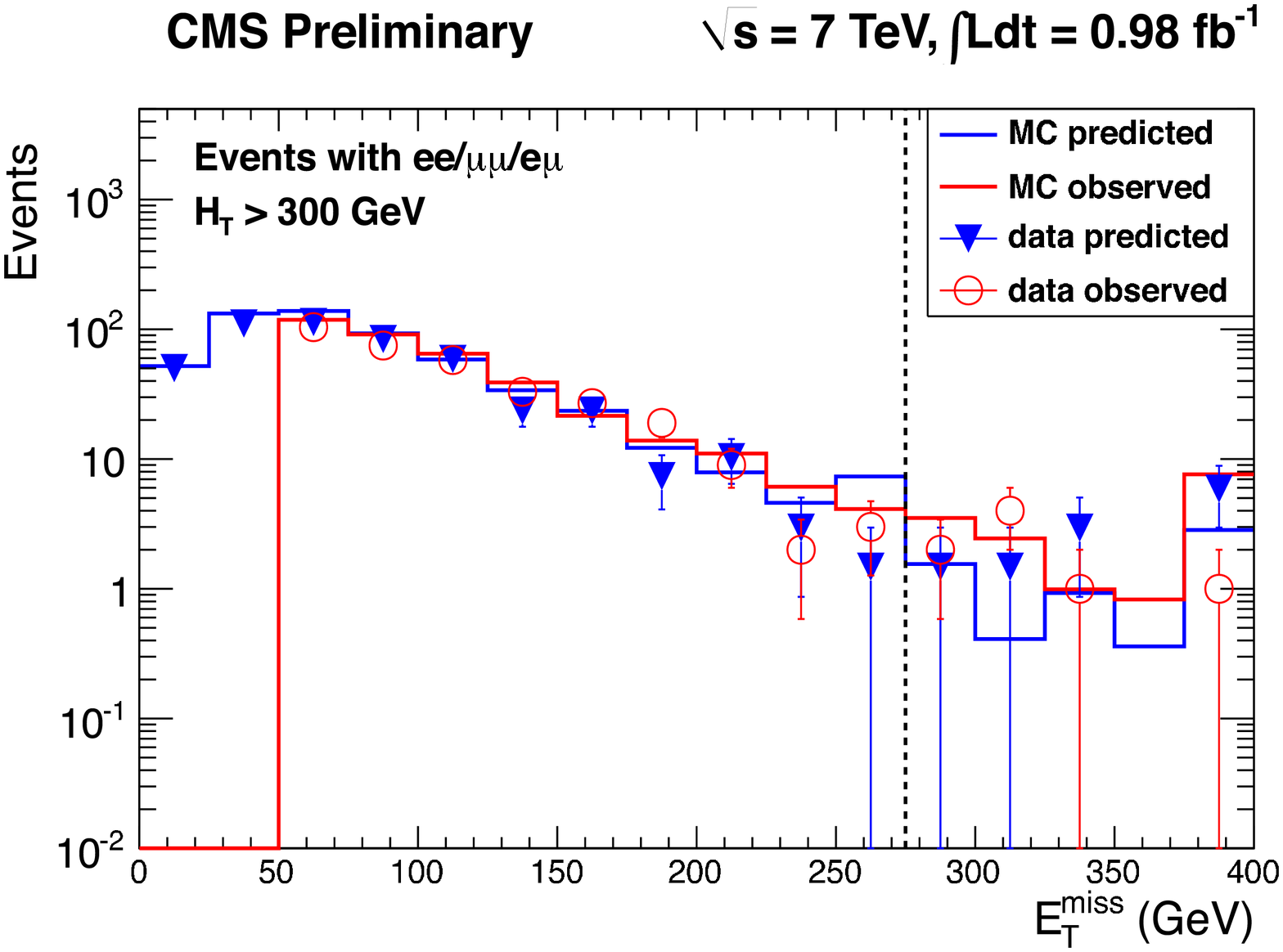}
\includegraphics[height=4cm]{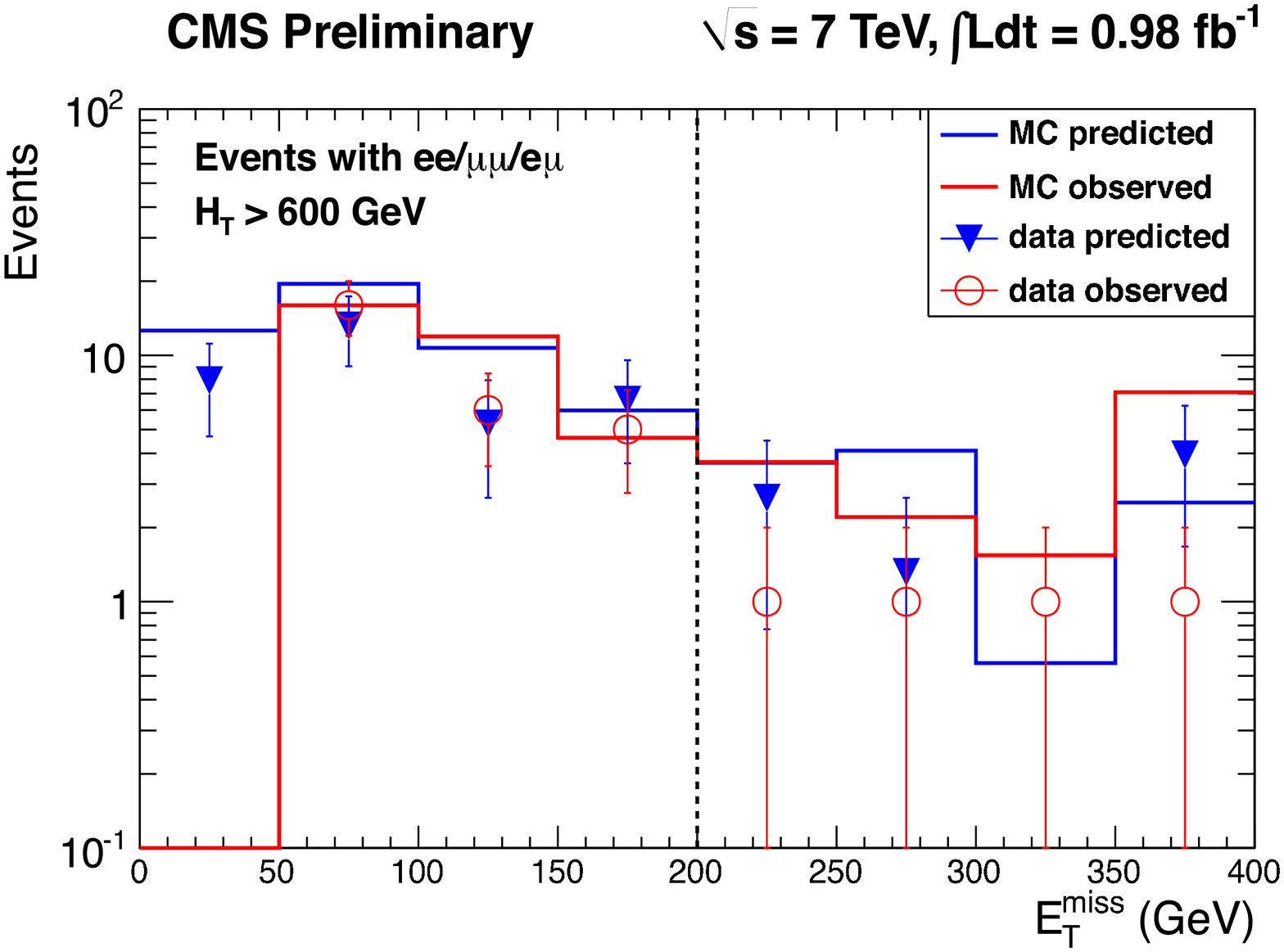}
\caption{Summary of background predictions and observed yields in the search regions for the $E^{Miss}_{T}$ distributions that signal regions are indicated by the vertical lines in the plots.}
\end{figure}

\begin{enumerate}
\item  High $E^{Miss}_{T}$ signal region: $H_T > 300$ GeV and $E^{Miss}_{T} > 275$ GeV
\item  High $H_{T}$ signal region: $H_T > 600$ GeV and $E^{Miss}_{T} > 200$ GeV
\end{enumerate}
Three independent data-driven estimation methods are used to perform counting experiments in these signal regions. For both signal regions, the observed yield is consistent with the predictions from MC and from the data-driven background estimates based on observed data (Figure $4$). It is concluded that no evidence for non-SM contributions to the signal regions is observed at L=$0.98 fb^{-1}$ data.

\section{Interpretation of Results and Outlook}

The results of the hadronic ($\alpha_{T}$) and leptonic (same-sign and opposite-sign) SUSY searches are interpreted in the context of cMSSM model \cite{cMSSM}. In the absence of a signal, limits on the allowed parameter space in the cMSSM were set which exceed those set by previous analyses. 

\begin{figure}[hbtp]
\centering
\includegraphics[height=4cm]{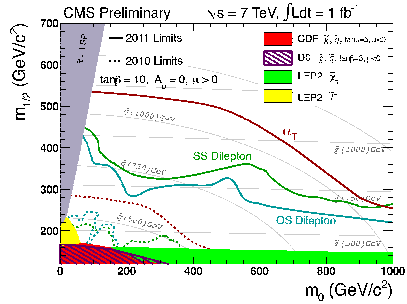}
\caption{Exclusion regions in the cMSSM corresponding to the observed limit from SUSY searches. The exclusion contours based on $34 pb^{-1}$ $2010$ data are also displayed.}
\end{figure}

No evidence for SUSY signature has yet been observed at L=$1.1 fb^{-1}$ data. The CMS collaboration expects to collect L=$5 fb^{-1}$ of data by the end of 2011.

\end{document}